 \documentclass[10pt,conference]{IEEEtran} 
\usepackage[keeplastbox]{flushend}
\ifCLASSINFOpdf
  \usepackage{graphicx}
  \usepackage{lipsum}
  \usepackage{caption}
  \usepackage{float}
  \usepackage{subfig}

\else
\fi
\IEEEoverridecommandlockouts
\usepackage{amsmath}

\usepackage[pscoord]{eso-pic}
\newcommand{\placetextbox}[3]{
	\setbox0=\hbox{#3}
	\AddToShipoutPictureFG*{
		\put(\LenToUnit{#1\paperwidth},\LenToUnit{#2\paperheight}){
			\vtop{{\null}\makebox[0pt][c]{#3}}}
	}
}
\placetextbox{.2}{0.055}{ 978-3-901882-98-2~\copyright~2017 IFIP}

\begin{document}
%
\title{Attacking Strategies and Temporal Analysis Involving Facebook Discussion Groups}




%
\author{\IEEEauthorblockN{Chun-Ming Lai\IEEEauthorrefmark{1},
Xiaoyun Wang\IEEEauthorrefmark{1},
Yunfeng Hong\IEEEauthorrefmark{1}, 
Yu-Cheng Lin\IEEEauthorrefmark{1},\\
S. Felix Wu\IEEEauthorrefmark{1},
Patrick McDaniel\IEEEauthorrefmark{2} and
Hasan Cam\IEEEauthorrefmark{3}
}
\IEEEauthorblockA{\IEEEauthorrefmark{1}
	University of California, Davis\\
\{cmlai,xiywang,yfhong,ycjlin,sfwu\}@ucdavis.edu
}
\IEEEauthorblockA{\IEEEauthorrefmark{2}
	Pennsylvania State University\\
	mcdaniel@cse.psu.edu
}

\IEEEauthorblockA{\IEEEauthorrefmark{3}
	U.S. Army Research Laboratory\\
	hasan.cam.civ@mail.mil
}
}

\maketitle

\begin{abstract}
Online social network (OSN) discussion groups are exerting significant effects on political dialogue. In the absence of access control mechanisms, any user can contribute to any OSN thread. Individuals can exploit this characteristic to execute targeted attacks, which increases the potential for subsequent malicious behaviors such as phishing and malware distribution. These kinds of actions will also disrupt bridges among the media, politicians, and their constituencies. 

For the concern of Security Management, blending malicious cyberattacks with online social interactions has introduced a brand new challenge. In this paper we describe our proposal for a novel approach to studying and understanding the strategies that attackers use to spread malicious URLs across Facebook discussion groups. We define and analyze problems tied to predicting the potential for attacks focused on threads created by news media organizations. We use a mix of macro static features and the micro dynamic evolution of posts and threads to identify likely targets with greater than 90\% accuracy. One of our secondary goals is to make such predictions within a short (10 minute) time frame. It is our hope that the data and analyses presented in this paper will support a better understanding of attacker strategies and footprints, thereby developing new system management methodologies in handing cyber attacks on social networks.
\end{abstract}


%
\IEEEpeerreviewmaketitle

\section{Introduction}
The number of people using online social networks (OSNs) such as Facebook and Twitter as their primary sources of news and information has grown dramatically over the past two decades. An important characteristic of OSNs is that they can be used to both receive and transmit (or share) information. Thus, anyone with Internet access can immediately type personal responses to official posts appearing on the White House Facebook page. This capability exposes OSNs to constant attacks in the form of false rumors, bias propagation, spam, and messages containing malicious URLs. Two factors make it difficult to build objective classifiers to reduce the potential for content-related attacks involving false rumors and spam: (1) message definition is difficult because all audiences have their own criteria (making it unsuitable for natural language processing), and (2) the labeling of large amounts of ground truth data is exceptionally expensive. Some researchers have tried detecting non-content-related attack events such as artificial retweets, but unlike content-related attack events, these are more likely to appear in advertisements and other commercial messages. In this paper we will focus on malicious URLs because of their potential to cause significant damage via OSNs. A large amount of cybercrime activity in the form of phishing, cyber-bullying, and online fraud is initiated by malicious URLs.

Handling cybercime is important in security management operations. Cybercrime --- especially that involving OSNs --- differs significantly from other forms of crime because of the skills that attackers use to hide behind multiple fake and compromised accounts. According to a long-held description of criminal acts known as routine activity theory (RAT) \cite{cohen1979social}, crimes entail three essential factors: a likely offender, a suitable target, and the absence of a capable guardian. As part of our attempt to apply RAT to malicious actions involving OSNs, we will try to determine what kinds of OSN environments are more likely to attract the attention of attackers, and what social activity patterns might be used to identify them. To accomplish these goals, we will focus on attacker motivations and incentives, using an idea from RAT to explain attacker strategies. Our assumption is that malicious URL attackers with limited resources tend to select targets that produce the biggest results. Accordingly, malicious comment timestamps are considered valuable tools for researchers to examine the activities of both attackers and defenders. Our proposed supervised learning framework is designed to identify malicious URLs, based on observations of significant differences in the evolving statuses of typically targeted and non-targeted OSN threads. We found that it is possible to predict, in near-real time, which threads are likely targets of attacks. Further, we analyzed the timestamps of malicious accounts that launch attacks, based on the known habits of attackers to spread the same or similar messages. 

We hope to make three contributions:
\begin{itemize}
	\item Present a novel method for accurately identifying post threads of interest to attackers, based on detailed evidence involving attacker intentions.
	\item Provide a detailed analysis of attackers in terms of their digital footprints and patterns, based on data gathered from more than 40,000 Facebook pages.
	\item Identify relative time intervals between consecutive attacks, and the absolute times when campaigns occur after initial posts are created. This information can be used to determine malicious campaign synchronization.
	
\end{itemize}
The rest of this paper is organized as follows. In Section \ref{two} we categorize OSN security threats and summarize past research on strategies for combating malicious behaviors targeting OSNs. In Section \ref{three} we describe our dataset and validation method, and present a formal problem statement. In Section \ref{four} we present and evaluate our proposed prediction framework based on the static and dynamic features of Facebook posts. In Section \ref{five} we present our results from a temporal analysis of malicious campaigns, and in Section \ref{six} we examine attacker identities based on previous behaviors. A conclusion is offered in Section \ref{seven}.

\section{Related Work} \label{two}
Previous researchers have studied two categories of OSN attack strategies. The first involves analyses of illegitimate content using lexical metadata such as average word length, average number of words per message (message length), ratio of uppercase to lowercase letters, and embedded URL features, among other factors. The primary goal in many of these studies is to distinguish spam messages from user-generated content  \cite{balakrishnan2016improving} \cite{gani2012towards} \cite{harsule2016n} \cite{grier2010spam}  \cite{stringhini2010detecting} \cite{rahman2012efficient}. With advancements in natural language processing, more abstract concepts such as message topic modeling  \cite{ma2016message} have been increasingly applied to build more complex language models. However, since these approaches usually employ blacklist filters or similar tools, they perform poorly when attackers change their context generation algorithms. The second group focuses on detecting fake and otherwise compromised accounts using anomaly detection algorithms that are either history profile-based or graph-based. Commonly used profile characteristics are age, image, description, number of followers, geolocation, and total number of posts. These data are used to construct classification systems and supervised learning algorithms aimed at identifying and blocking malicious accounts  \cite{alsaleh2014tsd} \cite{miller2014twitter} \cite{lee2010uncovering} \cite{rajadesingan2015sarcasm} \cite{huang2013analysis}. 

OSN connectivity and interaction features have been used to experiment with graph-based approaches to identifying accounts that exhibit inappropriate behaviors. Some researchers have observed that malicious accounts tend to have few connections with normal users but multiple connections with each other  \cite{cao2012aiding} \cite{danezis2009sybilinfer} \cite{viswanath2010analysis} \cite{zhu2012discovering} \cite{yu2006sybilguard} \cite{yu2008sybillimit}. However, to maximize the impact and shared effect in OSNs, many spam attacks are well-organized by a \emph{group} of accounts. Consequently, a novel detection method clustering a group of malicious accounts has been proposed recently. A GNOME panel applet called COMPA \cite{egele2013compa} detects compromised account by identifying similar changes in account behaviors within a short period of time. Two behavioral clustering approaches called CopyCatch \cite{beutel2013copycatch} and SynchroTrap \cite{cao2014uncovering} detect accounts that exhibit synchronized group patterns. Further, Ye \emph{et al.} \cite{ye2015discovering} have used group network footprints to detect opinion spammers. 
However, even though many efforts have been made to study attacker behaviors, few researchers have looked at intentions or acts of collusion involving two or more campaigns, especially in terms of the kinds of targets that attackers are interested in. One significant challenge to detecting the behaviors of attackers is their ability to quickly alter such behaviors in order to achieve their goals. Accordingly, there are more researchers attempting to study the characteristics of targets that attackers are interested in. Song \emph{et al.} \cite{song2015crowdtarget} have distinguished between tweets receiving retweets from crowdturfing accounts and tweets receiving retweets from normal accounts using standard retweet-based features. Cao \emph{et al.} \cite{cao2015detecting} have exploited post-based and click-based features to detect spam URLs in Twitter. In contrast, our aim is to use thread life cycle information (which is difficult for attackers to detect) to determine if and when attackers have an interest in specific Facebook discussion groups. To our knowledge, this is the first attempt to use a large sample of OSN news groups (with $>160$ million visitors) to perform a temporal analysis of OSN malicious campaigns as opposed to individual walls that may only attract small numbers of friends and acquaintances. 

\section{Data Description and problem statement} \label{three}
In this section, we first introduce Facebook as an online service of social media. After providing an overview scope of discussion groups dataset, we show that how we collect our data and how we validate the ground truth -- a comment contains malicious URLs.
\subsection{Facebook Discussion Groups}
Facebook is currently the world's most popular OSN worldwide. There are over 1.86 billion monthly active users which is a 17 percent increase year over year.\footnote{https://zephoria.com/top-15-valuable-facebook-statistics/} Typical of most OSNs, participants are encouraged to create individual accounts, manage their profiles, and build relationships by adding posts or making comments, sharing content, or expressing their approval of others’ content in the form of \emph{likes}. Besides functioning as an OSN, Facebook also plays an important role in online social media as a user-generated content platform for posting text, comments, digital photos, and videos --- the lifeblood of social media \cite{obar2015social}. 

Professional politicians and political organizations use Facebook to share announcements and information about what they believe to be interesting events in order to reduce distance from their constituencies. News organizations also rely on Facebook pages to encourage and support interactions with their audiences. Compared to ego networks, Facebook discussion groups are thought to resonate better with users who have weak ties with other users \cite{granovetter1973strength}. According to Granovetter \emph{et al.} \cite{granovetter1983strength}, individuals with few or weak ties tend to be deprived of information from distant social system locations, and are therefore confined to news from local sources and the views of close friends.
From attacker perspectives, the interpersonal networks that exist within and between Facebook discussion groups are more suitable than personal walls for launching attacks. The diffusion of rumors and other bits of information tend to be moderated by friend relationships, meaning that they are more likely to flow through groups of individuals with weak ties. Facebook discussion groups are much more popular than personal walls --- three examples are CNN, Fox News, and The White House. Since these and similar groups do not have privacy setting concerns, they are invaluable for researchers interested in conducting communication and cyber security studies. 

\begin{table*}
	\centering
	\caption{Data description}
	\label{datadesc}
	\begin{tabular}{|c|c|c|c|c|} \hline
		Page Name & Location &  Target Posts & Non-target  Posts & Total Comments \\ \hline
		\textbf{Middle East} & & & & \\
		Iraq News & Irbil, Iraq & 63 & 30,663 & 3,372,773 \\ 
		RassdNewsN & Cairo, Egypt & 18 & 48,559 & 4,819,892 \\ 
		Syrian Revolution & Aleppo, Syria & 213 & 187,690 & 2,667,453 \\ \hline
		\textbf{Asia} & & & & \\
		Taiwan Apples & Taipei, Taiwan & 110 & 32,364 & 2,175,145 \\ \hline
		\textbf{Europe} & & & & \\
		Le Monde & Paris, France & 119 & 14,822 & 1,058,524 \\ 
		BBC News & London, UK & 133 & 2,719 & 827,752 \\ \hline
		\textbf{US News} & & & & \\
		CNN & New York & 361 & 142,409 & 3,623,306 \\
		FOX & New York & 486 & 3,535 & 6,910,484 \\ \hline
		\textbf{US Politics} & & & & \\
		Barack Obama & D.C. & 1,297 & 2,110 & 10,647,232 \\
		The Obama White House & D.C. & 1,052 & 3,111 & 6,600,902 \\
		\hline\end{tabular}
\end{table*}

\subsection{Crawled Data}
Unlike tradition web crawler, OSNs crawler should consider coverage, real-time, scale and data quality. The most important thing is the re-crawling process since Facebook Discussion Groups are constantly receiving new posts and replies from hundreds of millions users around the clock. Our Data was crawled by an open source social crawler \footnote{https://github.com/dslfaithdev/SocialCrawler} SINCERE system \cite{erlandsson2015crawling} from 2011 to 2014. We focus on main news media pages in Asia, United States, Middle East and Europe. In the United States area, we also pay attention to two well-famous pages: Barack Obama and The White House. On each discussion page, each article and corresponding comments were collected in our database. Detailed information includes timestamps, Facebook account identification number and the raw texts of comments and articles. Totally we have 471,834 posts and 42,703,463 comments over ten pages on Facebook. The full dataset is described in Table \ref{datadesc}. The most important terms used in this study are defined as follows:

\begin{itemize}
	\item \textbf{Page}: a public discussion group
	\item \textbf{Original post}: an article posted on a Facebook discussion group
	\item \textbf{Comment}: articles written in response to an original post
	\item \textbf{Post thread}: the original post and all responses.
	\item \textbf{Malicious comment}: a comment whose raw text contains at least one malicious URL verified by our ground truth method.
	\item \textbf{Target post}: any thread containing at least one malicious comment.
	\item \textbf{Non-target post}: a post thread with no malicious comments.
	\item \textbf{Attackers}: a set of OSN users who leave comments, with at least one comment identified as malicious. 
	\item \textbf{Time series TS}: $TS_{created}$ indicates the precise time when an original post or comment is added to a page. $TS_{created}$($post$) indicates the time an original article is posted. $TS_j$ indicates a time period $j$ following the time of the original post, measured in minutes. $TS_{final}$ refers to the precise time 1 hour after the original post was created (i.e., \emph{final} = 60).
	\item \textbf{Number of comments} $N_{comment}$ (post,$TS_i$): the number of post comments collected at $TS_i$
	\item \textbf{Accumulated number of participants}
	$AccN_{comment}$ (post,$TS_i$): the number of post comments between $TS_i$ and $TS_{i-1}$.
\end{itemize}

	
	

\subsection{Ground truth Method and Problem statement}
In this section, we show that how do we evaluate our result and provide an official problem statement.
\subsubsection{Ground truth} ~\\ 
Our intention was to find a way to automatically identify and label malicious comments. The first step was to determine whether or not the raw text of a comment contained one or more URLs. When it did, we checked to see if it contained a redirect to a URL shortening service, and if so, we recovered the original URL. Next, each obtained URL was stored according to a key-value structure, in which the key was the URL domain, and the values included page identification number, post thread identification number, Facebook user identification number, and timestamp. Data were sorted according to individual keys.

We then used URLblacklists \footnote{http://urlblacklist.com/} to categorize the extracted URLs. This commercial service, which provides lists of specifically categorized websites to its subscribers, is included in web-filtering tools such as SquidGuard \footnote{http://www.squidguard.org}. In this study we considered 4 malicious URL domains from the 111 categories available to URLblacklists users: \emph{ads}: advert servers and banned URLs; \emph{malware}: websites known to contain malware, viruses, trojan horses, or backdoors; \emph{phishing}: sites known for attempting to trick users into giving out private information; and \emph{porn}: pornography. We also processed blacklisted URLs according to a key-value data structure in which keys represent individual URLs and values represent categories. Last, for the two sorted key-value hashmaps we used two indexes (starting from the first of each sorted list), where time complexity is $O(log(mn))$, with $m$ representing the number of URLs in our dataset and $n$ the number of blacklisted URLs. Our experiment used more than 5M comments with URLs and 1M blacklisted URLs, indicating a need for an efficient algorithm to perform labeling on a large scale.

\subsubsection{Problem statement}  ~\\
As stated above, our goal is to predict which posts are more likely to attract (and subsequently spread) malicious URLs. In other words, for any post thread $p$ in a social media platform discussion group, the main target detection problem is predicting whether $p$ contains at least one malicious comment via a classifier --- $c \to $\{$target, nontarget$\} --- based on one set of $F_{macro}$ macro static post features and one set of $F_{micro}$ micro dynamic features. Since neither $F_{ macro}$ nor $F_{micro}$ are related to content, they are difficult for attackers to comprehend. In the following section we will describe in detail their features and our proposal for a classifier. 



\section{ Features and Prediction} \label{four}
Since attackers are likely to attempt to spread malicious URLs during periods of peak discussion activity in order to influence the largest potential audience, a useful strategy for studying attacker behavior is identifying malicious comments across a series of discussion threads. In this section we will examine the characteristics of target and non-target post threads in terms of \emph{popularity} and \emph{evolution} over time. This serves as the basis for our prediction framework and evaluation procedure.

\subsection{Macro static features}
Identifying OSN content that is likely to attract traffic and user interaction is a core issue for researchers. Two concepts from the fields of communications and psychology --- the bandwagon effect \cite{wang2015bandwagon} and information cascades \cite{bikhchandani1992theory}  --- have been used to explain article popularity and duration. We will address popularity features first. 

\begin{enumerate}
	\item \textit{Spanning Time:}
	OSNs use recommendation systems to promote information on homepages. Every post thread has its own life cycle, although determining how each OSN implements its respective life cycle is a challenging task. We observed that targeted post threads tend to have longer survival times. In this study, post thread span time was defined as the number of days from the creation of a timestamp for a post thread to the timestamp of its last received comment.
	
	\item \textit{Number of Comments:}
	Anyone who has spent time reviewing an OSN has observed that some post threads have large numbers of comments while others have none or very few. The total number of responses to a post can serve as a simple indicator of post thread popularity.
	
	\item \textit{Number of Participants:} Since individual users can post multiple responses in a thread, care must be taken to determine the precise number of participants. 
	
	\item \textit{Number of Likes regarding posts:}  
	It is important to remember that it takes much less effort to register a “like” than it is to write and post a response.
	
	\item \textit{Number of Likes regarding all corresponding comments}: This metric reflects the number of users who agree with individual comments rather than the original post.
	
\end{enumerate}

\begin{figure*}[]
	\hspace*{-0.5cm} 
	\includegraphics[scale=0.6]{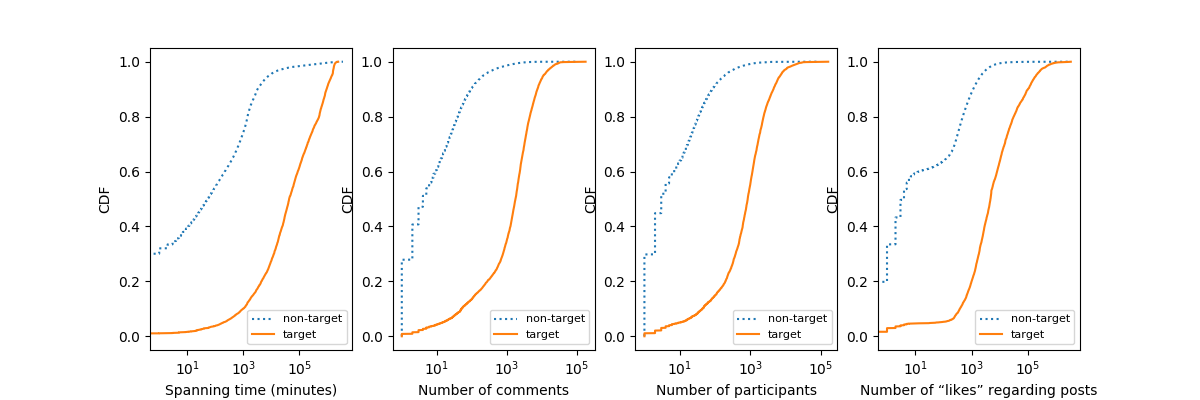}
	\caption{Macro Static Features CDFs between targets and non-target posts}
	\label{global_CDF}
\end{figure*}

Figure \ref{global_CDF} shows data for cumulative distribution function (CDF) macro static features between target and non-target posts. As shown, target post threads had longer duration times; for non-target posts the duration times were less than 10 minutes. According to Figure \ref{global_CDF} , posts that are likely targets tend to attract more likes, comments, and users than non-target posts --- evidence that attackers make the necessary effort to find popular threads.

\subsection{Micro dynamic evolving life time}
The literature contains few studies focused on thread duration. We believe that longer durations, longer discussions, and larger numbers of opinions should be expected for threads discussing political news, for subjective announcement threads (e.g. ObamaCare), and threads aimed at more heterogeneous audiences. When attempting to identify correlations (if any) between post thread participation and elapsed time, Wang \emph{et al.} \cite{wang2015bandwagon} found that the first 15 minutes are critical for thread development. Castillo \emph{et al.} \cite{castillo2014characterizing} improved the timeliness aspect of predictive models for news article total visits and shelf life by incorporating social media reactions. We used a similar idea to clarify whether post threads are “promoted” to attackers by OSN recommendation systems --- that is, we analyzed temporal thread development using the variables $t_{window}$ and $t_{final}$.

To establish a definition for {discussion atmosphere vector } (DAV), we first used the definition of accumulated number of participants given in section 3.2, using 5 minutes for the $i$ value and 1 hour for the $t_{final}$ value. This resulted in the following equation: 

\begin{equation}
\begin{aligned}
DAV(Post)_{t_n} = [AccNcomment(Post,t_1),\\
 AccNcomment(Post,t_2), ...,  \\ 
AccNcomment(Post,t_n)]
\end{aligned}
\end{equation}

\begin{figure}
	\centering
	\includegraphics[width=.5\columnwidth]{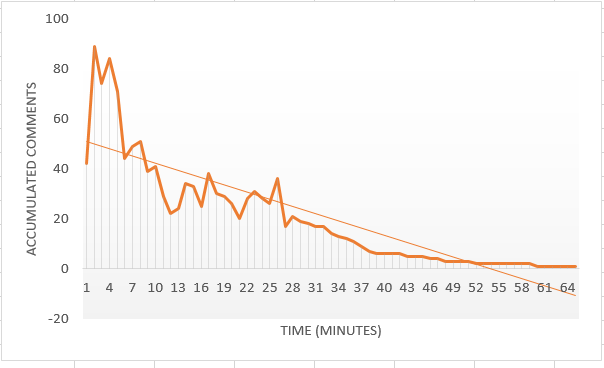}
	\caption{Classical post decay with respect to time}
	\label{post_decay}
\end{figure}

In the example shown in Figure \ref{post_decay}, for the initial $t_i$, $AccNcomment(Post,t_i)$ increases slightly, peaks, and experiences a long-tail decay. We then used DAV to classify target and non-target post threads based on dynamically committed comments with respect to time, since each DAV element represents the number of comments accumulated in each $t_i$.  

\begin{figure}
	\centering
	\includegraphics[width=.9\columnwidth,height=.5\linewidth]{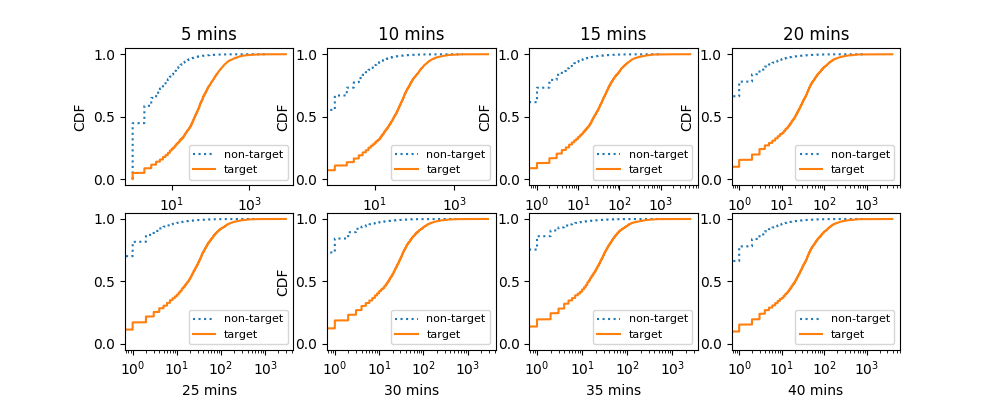}
	\caption{Post DAVs between target and non-target posts during the first 40 minutes following initial posts after posts created}
	\label{evolving}
\end{figure}
Figure \ref{evolving} presents CDF data for target and non-target posts for $DAV_{40mins}$ with a time window of 5 minutes. As shown, in most cases target posts have larger numbers of comments than non-target posts for each time slot. 

\subsection{Evaluation}
In the real world dataset described in Table \ref{datadesc}, we noted an obvious imbalance between target and non-target posts, and therefore used Synthetic Minority Over-sampling Technique (SMOTE) \cite{chawla2002smote} to balance the dataset. We then normalized both macro static and micro dynamic features so that their values were between 0 and 1, using 75\% of the data for training purposes and saving the rest for testing three popular machine learning classifiers, Naive Bayes, Adaboost and Decision Tree from the scikit-learn library \cite{scikit-learn}, and we use the evaluation tools introduced in Section \ref{three}. 

\begin{figure}
	\centering
	\includegraphics[width=1.0\columnwidth]{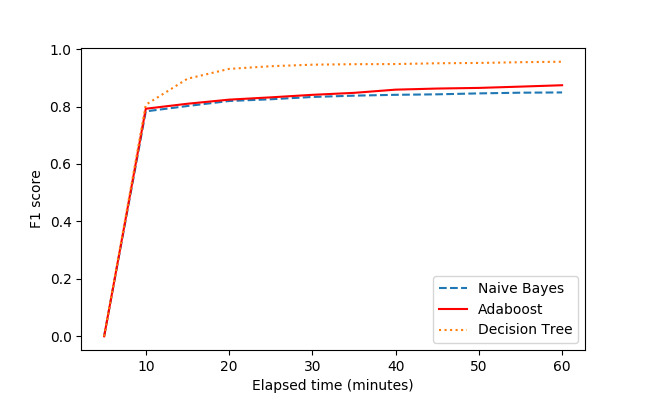}
	\caption{F1 score with respect to $t_{final} = 60$ $minutes$}
	\label{by_time}
\end{figure}

Table \ref{evaluation} presents classification results for macro static and micro dynamic features, respectively. Our results indicate good robustness for the micro dynamic feature in predicting target posts with low false positive rates. We then attempted to determine the optimum time for collecting comments for each time window --- that is, the best value for $t_{final}$. According to the data shown in Figure \ref{by_time}, it took ten minutes to achieve an F1-score of approximately 0.8. While a large number of time slots would definitely improve these results, it is important to consider the consequent increase in time required to train the model.

\begin{table}
	\caption{Classifier evaluation data}
	\label{evaluation}
	\begin{tabular}{c|c|c|c} \hline
		Classifier & Precision &  Recall & F1 score \\ \hline
		\textbf{Macro static features} & & &  \\
		Naive Bayes& 0.72 & 0.68 & 0.66 \\ 
		Adaboost & 0.77 & 0.76 & 0.76  \\ 
		Decision Tree & 0.95 & 0.95 & 0.95  \\ \hline
		
		\textbf{Micro Dynamic Features} ($1$ hour) & & &  \\
		Naive Bayes& 0.85 & 0.85 & 0.85 \\ 
		Adaboost & 0.87 & 0.87 & 0.87  \\ 
		Decision Tree & 0.96 & 0.96 & 0.96  \\ \hline
		
		\textbf{Mixed Features} & & &  \\
		Naive Bayes& 0.78 & 0.73 & 0.72 \\ 
		Adaboost & 0.86 & 0.86 & 0.86  \\ 
		Decision Tree & 0.98 & 0.98 & 0.98  \\ \hline
		\hline\end{tabular}
\end{table}

\section{Temporal Analysis of Malicious Campaigns} \label{five}
After selecting appropriate target posts, we turned our attention to \emph{temporal} considerations. We used several temporal variables to analyze attacker strategies.
A temporal analysis of detected malicious URLs was required, based on our observation that attack timestamps can serve as useful digital evidence for forensic purposes. After identifying targets based on the DAV and popularity features described in section \ref{four}, we tried to determine the most likely thread stage for attacks (early, middle or late), and whether attacks tend to occur within specific time periods or if they are randomly distributed. We examined these questions across different regions and different types of malicious URLs.

\subsection{Hotspot Crime time}
In a typical OSN scenario, initial posts are followed by a period of concentrated discussion, and as time progresses, intervals between consecutive comments increase. Since attackers want to reach the largest number of thread participants, it seems reasonable to assume that they will benefit the most by attacking during early thread stages. However, they might want to take advantage of the Facebook function that notifies all previous commenters regarding new content, and wait until a later thread stage so that all previous commenters are exposed to the malicious URL. 

\begin{figure}
	\includegraphics[height=4.7cm,width=1.0\columnwidth]{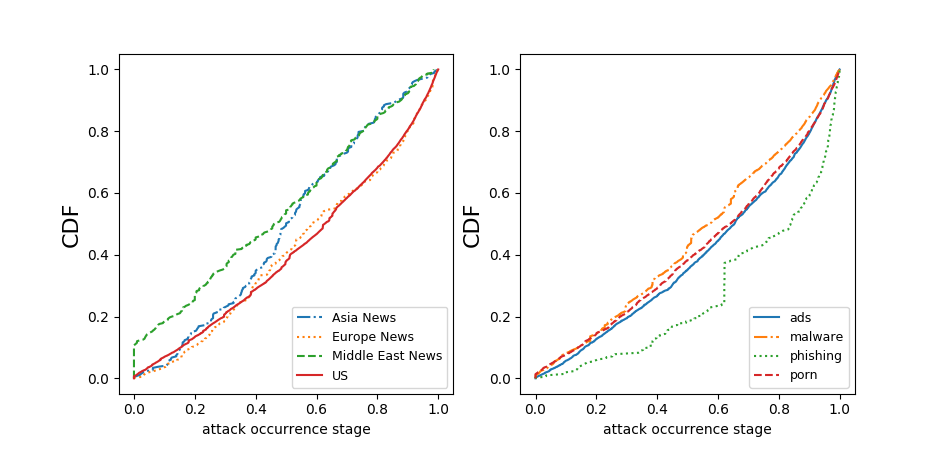}
	\caption{CDFs of malicious campaign occurrences plotted against life stages of targeted threads.}
	\label{position}
\end{figure}

To determine attacker preferences, we sorted all comments by their comments’ timestamps. We then examined the occurrence stages of malicious URLs. Results are shown in Figure \ref{position}. The X-axis represents the relative position of a post, with 0 indicating the beginning of the thread and 1 the end. The left part indicates that attackers who focused on Middle Eastern and Asian news threads tended to launch their attacks during early stages. Approximately 65\% of attacks focused on European and American news threads occurred during later stages. The right part indicates that for the same countries, phishing attacks were more likely than malware, advertising, and porn-focused attacks to be launched during late thread stages.

\subsection{Exact time after posts have been created}
\begin{figure}
	\includegraphics[height=4.7cm,width=1.0\columnwidth]{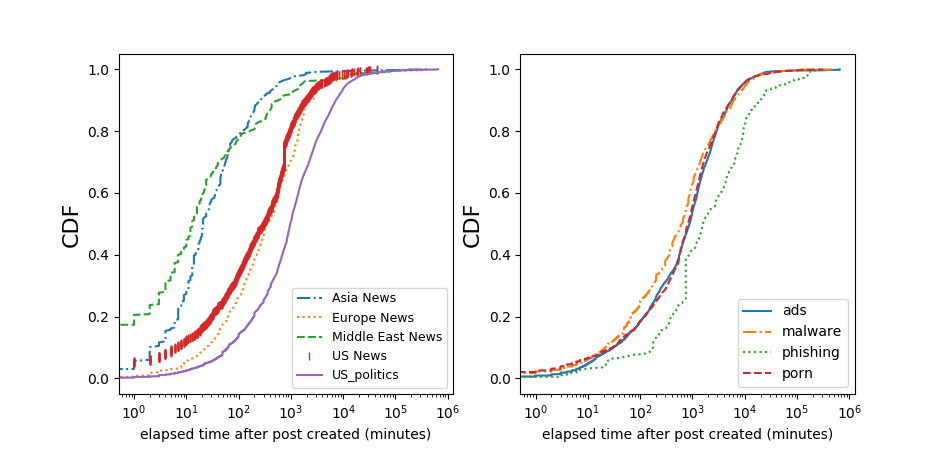}
	\caption{CDFs of malicious campaign occurrences plotted against amounts of time passed since the posting of initial target thread posts.}
	\label{m_afterpost}
\end{figure}

We attempted to determine the precise time intervals of malicious campaigns --- specifically, time intervals between malicious URL timestamps and the dates/times that targeted threads were created. Figure \ref{m_afterpost} presents our results in terms of geographic area and attack type. We found that Asian and Middle East news-oriented pages had shorter response times, with approximately $75\%$ of all attacks launched within 100 minutes of initial article posts, compared to only $30\%$ for European and American news-oriented pages. We were impressed by our finding that nearly one-half of all attacks were launched within one day --- for Asian and Middle East-oriented threads the rate exceeded $90\%$. We noticed that approximately one-half of all attacks worldwide were launched within one day.

\subsection{Time interval after last attack on the same page}
\begin{figure}
	\centering
	\includegraphics[height=4.7cm,width=1.0\columnwidth]{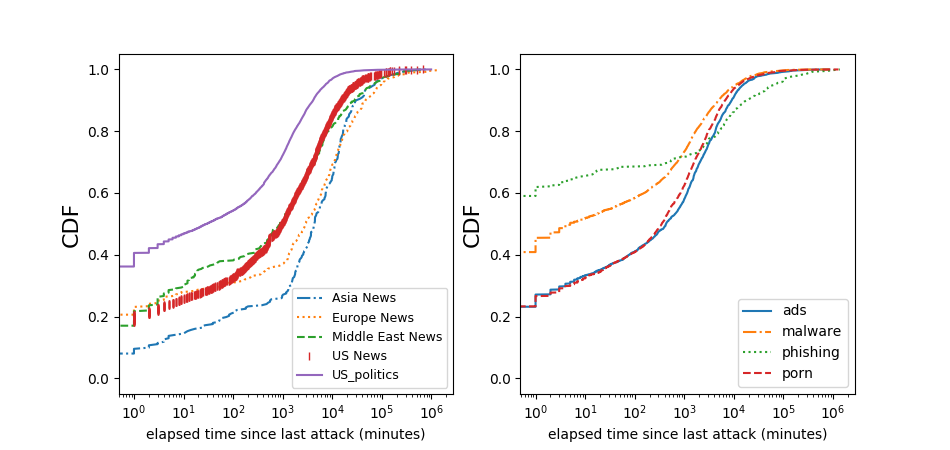}
	\caption{CDFs of malicious campaign occurrences plotted against amounts of time passed since the most recent prior attack on the same page.}
	\label{m_afterlast}
\end{figure}

Attackers are motivated to distribute malicious URLs to as many individuals as possible, via single or multiple accounts. We believe that multiple accounts are more harmful because such attacks can create false impressions that many people are responding within a short time period. In order to determine when attacks are synchronized or distributed, we examined time intervals following preceding malicious URL attacks. The data in Figure \ref{m_afterlast} show that attacks on pages discussing US politics tend to be concentrated within short time periods. Thus, when a malicious URL is detected, there is greater than $40\%$ probability that another malicious URL will appear within 10 minutes, compared to $20\%$ probability for other pages. Regarding type, we found that malware and phishing attacks tend to occur within shorter time periods compared to attacks featuring advertisements and pornography.

\subsection{Occurrence by date}
\begin{figure}
	\includegraphics[height=4.7cm,width=.9\columnwidth]{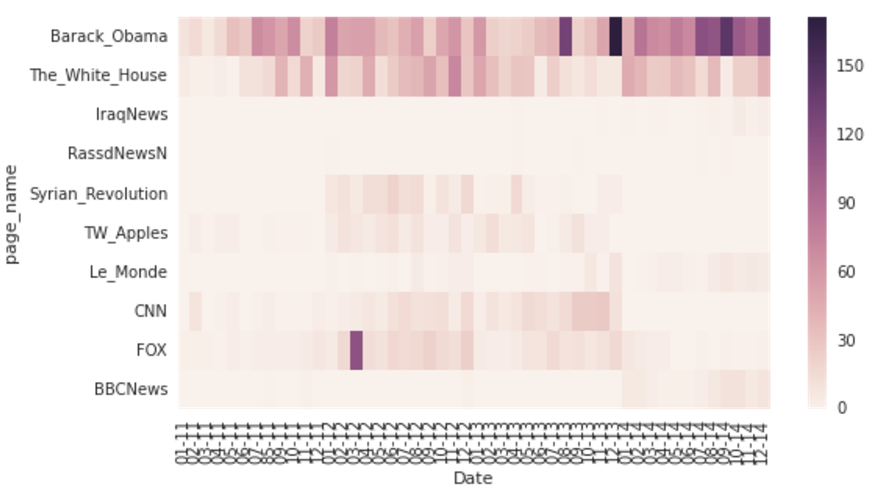}
	\caption{Heatmap evidence showing that threads in the sample that addressed American political issues attracted more attacks than other topics, and that threads posted in 2014 attracted more attacks than threads posted in 2011.}
	\label{heat_map_by_date}
\end{figure}

We are also interested in the hotspot time for attackers to launch malicious campaigns. Reports\footnote{https://www.washingtonpost.com/business/economy/russian-propaganda-effort-helped-spread-fake-news-during-election-experts-say/} show that there exists net army to spread fake news during election, therefore we assume that malicious URLs campaigns would focus on certain time such as the date before election or after a scandal. Figure \ref{heat_map_by_date} shows the heatmap of malicious URLs distributed across ten public pages and from 2011 to 2014, separated by month. We observe that the number of malicious URLs on US politics are greater than others, also 2014 is more severe than other years. 


\section{Characterize Malicious Accounts} \label{six}
In terms of social capital, OSN accounts with longer lifespans are more valuable than those with shorter lifespans and fake accounts. In this section we will discuss the details of malicious messages and attackers as part of our attempt to understand their synchronous group intentions.

\subsection{Accounts Macro-features}
\begin{figure}
	\centering
	\includegraphics[height=7cm,width=1.0\columnwidth]{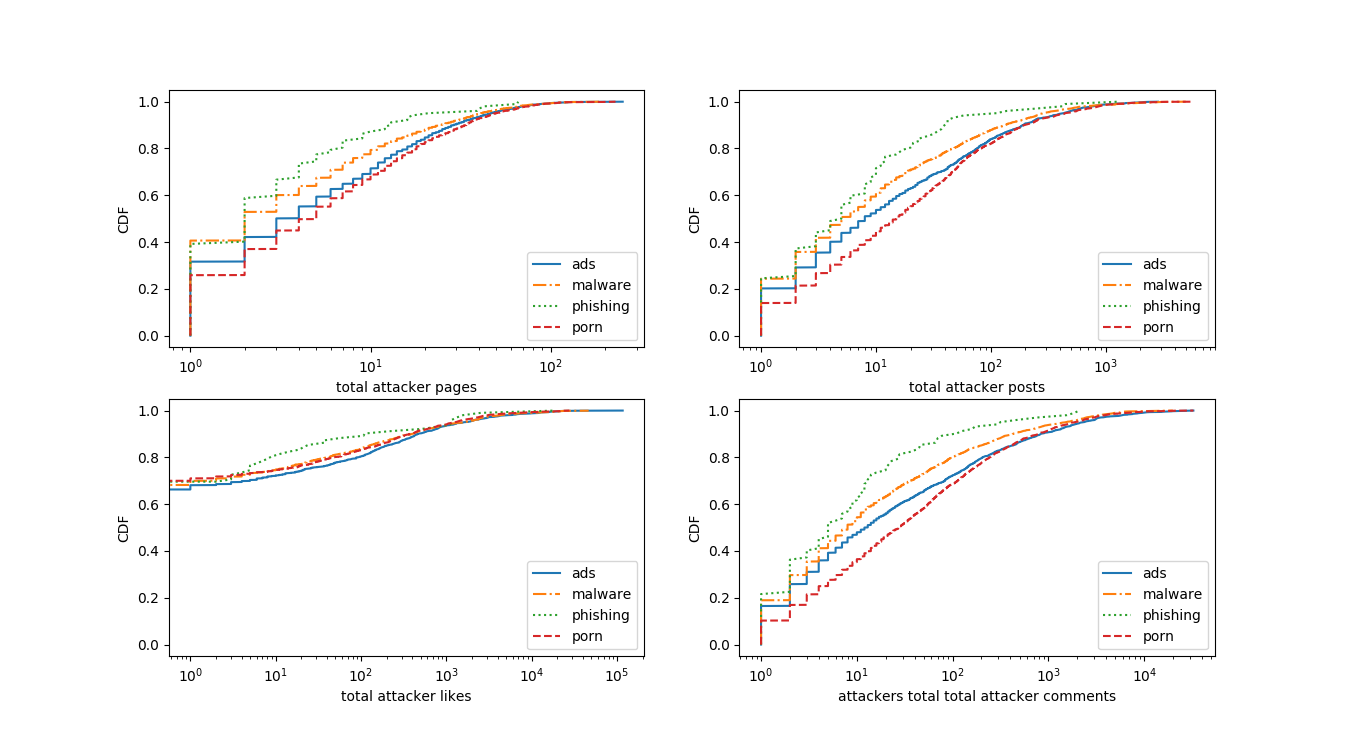}
	\caption{CDFs of the macro digital footprints of thread attackers.
}
	\label{global}
\end{figure}

In individual discussion threads it is easy to determine who supports an idea and who does not. However, there is still value in taking a user-oriented view to determine whether specific users are consistent --- that is, do they always commit malicious campaigns, or are they sometimes posting non-malicious content? Are they spreading the same messages to different targets? Are they only focusing on politics-related pages? To answer these questions we analyzed data gathered from more than 40,000 public Facebook pages between 2011 and 2016. Figure \ref{global} presents data for the macro-features of those pages, including total likes and comments from all 40,000 pages. Most malicious accounts were fake, and therefore had less activity --- over $70\%$ of those we identified had zero “likes,” which is unusual in light of how easy it is to express a “like” on an OSN. We found that phishing-focused accounts tended to have fewer pages, posts, likes, and comments compared to other types. 

\subsection{Account Digital Footprint}
Except the Macro-features we described, we are also interested in each activity of malicious account. Here, we compare history behavioral patterns and message similarity between malicious accounts and sampled normal accounts. We sampled  10,000 normal users, 1,000 from each page in Table \ref{datadesc}. Our data include more than 40,000 public pages around the world from 2011 to 2016.

\begin{figure}
	\centering
	\includegraphics[width=.9\columnwidth]{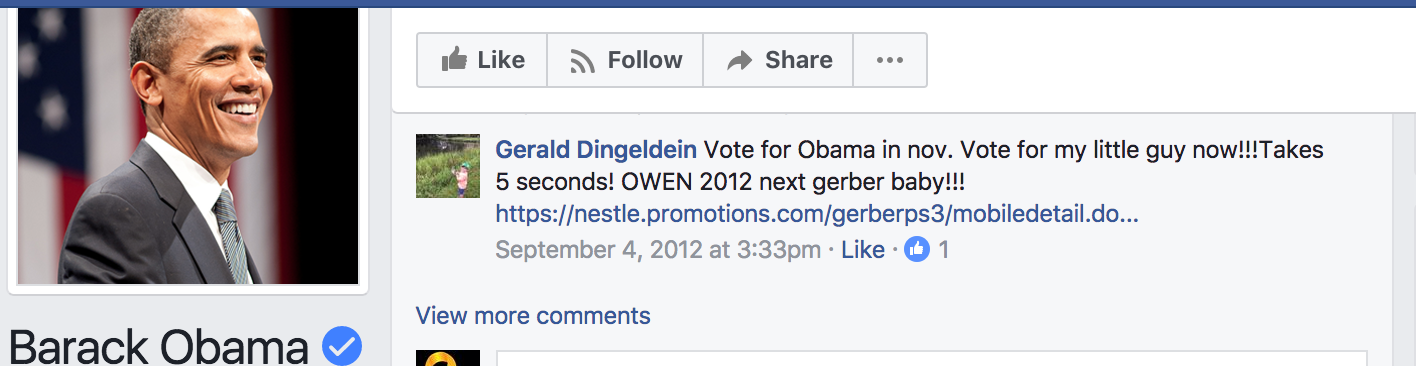}
	\caption{An example of detected malicious messages posted at least 10 times by the same account holder.}
	\label{example}
\end{figure}

\subsubsection{Response time to post thread}
\begin{figure}
	\includegraphics[height=4.7cm,width=.9\columnwidth]{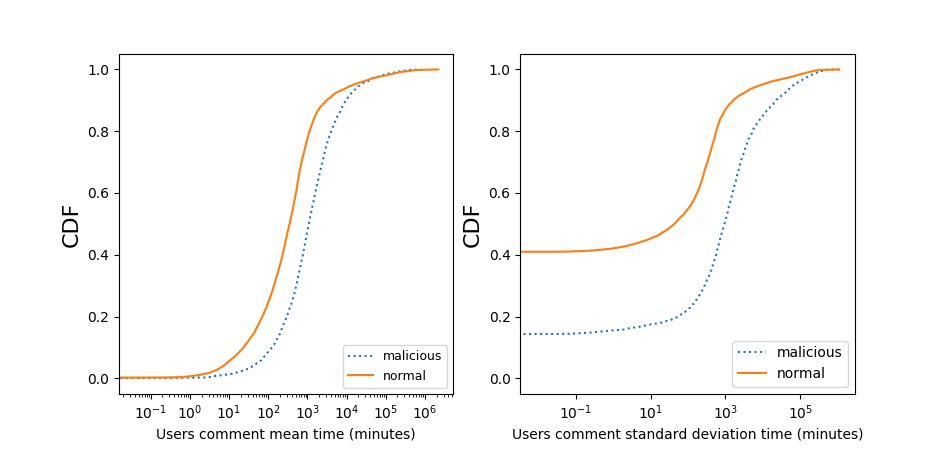}
	\caption{CDFs for mean numbers of user comments and their standard deviations plotted against amounts of time passed since the posting of initial posts. 
}
	\label{user_response_time}
\end{figure}

Our hypothesis is that if ones objective is attempt to launch malicious campaigns to public, they may be usually online and wait for target posts. Once the post thread starts, those malicious accounts would be the first few users to lead the discussion. Second strategy is that this duplicate messages has been propagated 10 times by one account (Figure \ref{example}), most of them appeared in the early stage of targets posts, the commenting time vector of this user = $[6194,5650,1,8,9,11,12,13,14,18]$. Here we examine the mean and standard deviation between malicious accounts and normal accounts, the result is as Figure \ref{user_response_time}. Obviously, malicious accounts tends to be active in the later stage of post threads compared with normal accounts. Moreover, we found that the standard deviation of malicious account is apparently greater than normal accounts, which refers that legitimate users usually commit after a fixed time interval after a post has been created.

\subsubsection{Same content, different accounts}
\begin{figure}
	\centering
	\includegraphics[width=.9\columnwidth]{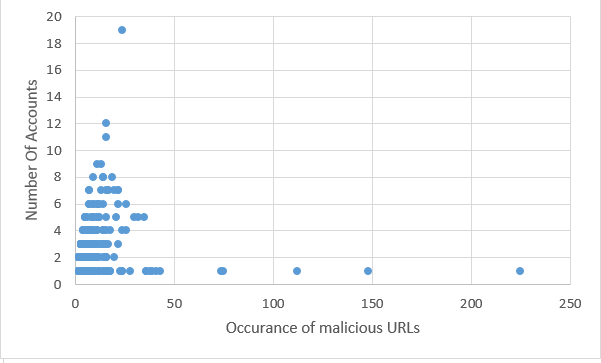}
	\caption{Total number of accounts in the sample plotted against numbers of
		occurrences of malicious URL posts}
	\label{XYscattern}
\end{figure}

We clustered the exactly same malicious URLs as one malicious campaign, then we are interested in each campaign, how many accounts would be used to perform the campaign? Figure \ref{XYscattern} shows that normally less than 10 accounts spread one malicious URLs, most of them are distributed and not related. However, there are two types campaigns needed to be considered. (1) many accounts, one malicious URL (upper left corner of Fig. \ref{XYscattern}): We assume this type of attacks are the most harmful on OSNs since there must be someone behind the monitor and control these accounts to spread the same URL.
(2) One account, many copies of malicious URL (lower right corner of Fig. \ref{XYscattern}): The other situation is that one account spread as many copies of malicious URL as they can.

\section{Conclusion} \label{seven}

We describe our work on detecting and characterizing large-scale malicious URLs campaigns using public Facebook Discussion Groups dataset.  The dataset includes news media in Facebook across main regions around the world between 2011 and 2014. Totally we examine 42,703,463 comments, and label them in an efficient way using a third party service.

Motivated by Routine Activity Theory and Information Cascade, we present an interdisciplinary research combining with computer science, journalism and criminology. We then propose a novel target detection method using macro static features regarding popularity and micro dynamic evolving lifetime in a near real-time way. Through the result, we perform an in-depth analysis on attackers strategies. We analyze the hotspot crime time regarding different countries and different types of campaigns. We also calculate the occurrence time (1) by exact date, (2) after posts have been posted and (3) after last campaigns. Our result not only offer a recommendation system for attackers to launch a relatively effective campaign but also help defenders to inspect suspected targets with limited resource.

The initial result we obtained provides us encouraging new insight regarding cyber attacks leveraging online social interactions. Even though we believe that there are many other strategies to study for future work, our current observation enables us to leverage these techniques to develop new response strategies in handling such attacks.






\section*{Acknowledgment}

The effort described in this article was partially sponsored by the U.S. Army Research Laboratory Cyber Security Collaborative Research Alliance under Contract Number W911NF-13-2-0045. The views and conclusions contained in this document are those of the authors, and should not be interpreted as representing the official policies, either expressed or implied, of the Army Research Laboratory or the U.S. Government. The U.S. Government is authorized to reproduce and distribute reprints for Government purposes notwithstanding any copyright notation hereon.



%
{\bibliographystyle{IEEEtran}
	\bibliography{reference.bib}}

\end{document}